\documentclass[prd,aps,floatfix,nofootinbib,11 pt]{revtex4}
\usepackage{amssymb}
\usepackage{amsmath,graphicx,color,epsfig}

\input{tcilatex}

\begin{document}

\title{Decoherence and entanglement degradation of a qubit-qutrit system in
non-inertial frames}
\author{M. Ramzan\thanks{%
mramzan@phys.qau.edu.pk} and M. K. Khan}

\address{Department of Physics Quaid-i-Azam University \\
Islamabad 45320, Pakistan}


\begin{abstract}
We study the effect of decoherence on a qubit-qutrit system under the
influence of global, local and multilocal decoherence in non-inertial
frames. We show that the entanglement sudden death can be avoided in
non-inertial frames in the presence of amplitude damping, depolarizing and
phase damping channels. However, degradation of entanglement is seen due to
Unruh effect. It is shown that for lower level of decoherence, the
depolarizing channel degrades the entanglement more heavily as compared to
the amplitude damping and phase damping channels. However, for higher values
of decoherence parameters, amplitude damping channel heavily degrades the
entanglement of the hybrid system. Further more, no ESD is seen for any
value of Rob's acceleration.\newline
\end{abstract}

\pacs{04.70.Dy; 03.65.Ud; 03.67.Mn}
\maketitle

Keywords: Quantum decoherence; entanglement; non-inertial frames.\newline

\vspace*{1.0cm}

\vspace*{1.0cm}



\section{Introduction}

Quantum information and quantum computation can process multiple tasks which
are intractable with classical technologies. Quantum entanglement is no
doubt a fundamental resource for a variety of quantum information processing
tasks, such as super-dense coding, quantum cryptography and quantum error
correction [1-4]. Strongly entangled, multi-partite states of qubits and
qutrits are a central resource of quantum information science. These are
frequently used in constructing many protocols, such as teleportation [5],
key distribution and quantum computation [6]. The phenomenon of sudden loss
of entanglement also termed as \textquotedblright entanglement sudden death"
ESD has been investigated by a number of authors for bipartite and
multipartite systems [7-10]. Yu and Eberly [11, 12] showed that entanglement
loss occurs in a finite time under the action of pure vacuum noise in a
bipartite state of qubits. They found that, even though it takes infinite
time to complete decoherence locally, the global entanglement may be lost in
finite time.

Recently, researchers have focused on relativistic quantum information in
the filed of quantum information science due to conceptual and experimental
reasons. In the last few years, much attention had been given to the study
of entanglement shared between inertial and non-inertial observers by
discussing how the Unruh or Hawking effect will influence the degree of
entanglement [13--22]. Most of the investigations in non-inertial frames are
focused on the study of the quantum information in bipartite qubit system
with one subsystem as an accelerated. Since, it is\ not possible to
completely isolate a quantum system from its environment. Therefore, one
needs to investigate the behavior of entanglement in the presence of
environmental effects. A major problem of quantum communication is to
faithfully transmit unknown quantum states through a noisy quantum channel.
When quantum information is sent through a channel, the carriers of the
information interact with the channel and get entangled with its many
degrees of freedom. This gives rise to the phenomenon of decoherence on the
state space of the information carriers. Implementation of decoherence in
non-inertial frames have been investigated by different authors [23, 24].
Peres-Horodecki [25, 26] have studied entanglement of qubit-qubit and
qubit-qutrit states and established separability criterion. According to
this criterion, the partial transpose of a separable density matrix must
have non-negative eigenvalues, where the partial transpose is taken over the
smaller subsystem for qubit-qutrit case. For nonseparable states, the sum of
the absolute values of the negative eigenvalues of the partial transpose
gives the degree of entanglement of a density matrix also termed as
negativity. Ann et al. [27] have studied a qubit-qutrit system where they
have shown the existence of ESD under the influence of dephasing noise.

In this paper, we study the effect of decoherence on a qubit-qutrit system
in non-inertial frames by considering different noise models, such as,
amplitude damping, depolarizing and phase damping channels. We consider
different couplings of the system and the environment where the system is
influenced by global, local, or multi-local noises modeled in a number of
scenarios. We show that entanglement degradation occurs for various coupling
of the system and the environment. It is shown that different environments
degrade the entanglement of the hybrid system differently.

\section{Qubit-qutrit system in non-inertial frames}

We consider a composite system of a qubit $A$ and a qutrit $B$ that is
coupled to a noisy environment both collectively and individually. Local and
multi-local couplings describe the situation when the qubit and qutrit are
independently influenced by their individual noisy environments. Whereas,
the global decoherence corresponds to the situation when it is influenced by
both collective and multilocal noises at the same time. The term collective
coupling means when both the qubit and qutrit are influenced by the same
noise. The state shared by the two parties is an entangled qubit-qutrit
state of the form [28]%
\begin{equation}
\rho _{AR}=\frac{1}{2}\left. \left(
\begin{array}{c}
\cos ^{2}r(|01\rangle _{AR}\left\langle 01\right\vert +|01\rangle
_{AR}\left\langle 10\right\vert +|10\rangle _{AR}\left\langle 01\right\vert
+|10\rangle _{AR}\left\langle 10\right\vert ) \\
+\sin ^{2}r(|02\rangle _{AR}\left\langle 02\right\vert +|12\rangle
_{AR}\left\langle 12\right\vert )%
\end{array}%
\right) \right.
\end{equation}%
where the two modes of Minkowski spacetime that correspond to Alice and Rob
are $|\eta \rangle _{A}$ and $|\eta \rangle _{R}$ respectively. We assume
that Alice remain stationary while Rob moves with uniform acceleration. It
is important to mention here that the above state is obtained after taking
the trace over unobserved region IV [28]. The interaction between the system
and its environment introduces the decoherence to the system, which is a
process of the undesired correlation between the system and the environment.
The evolution of a state of a quantum system in a noisy environment can be
described by the super-operator $\Phi $ in the Kraus operator representation
as [29]

\begin{equation}
\rho _{f}=\Phi \rho _{i}=\sum_{k}E_{k}\rho _{i}E_{k}^{\dag }  \label{E5}
\end{equation}%
where the Kraus operators $E_{i}$ satisfy the following completeness relation

\begin{equation}
\sum_{k}E_{k}^{\dag }E_{k}=I  \label{5}
\end{equation}%
We have constructed the Kraus operators for the evolution of the composite
system from the single qubit Kraus operators (see table1 ) and qutrit Kraus
operators as given in equations (5-8) by taking their tensor product over
all $n\otimes m$ combination of $\pi \left( i\right) $ indices

\begin{equation}
E_{k}=\underset{\pi }{\otimes }e_{\pi \left( i\right) }  \label{6}
\end{equation}%
where $n$ and $m$ correspond to the number of Kraus operators for a single
qubit and qutrit channel respectively. The single qutrit Kraus operators for
the amplitude damping channel are given by [30]

\begin{equation}
E_{0}=\left(
\begin{array}{ccc}
1 & 0 & 0 \\
0 & \sqrt{1-p} & 0 \\
0 & 0 & \sqrt{1-p}%
\end{array}%
\right) ,\ \ E_{1}=\left(
\begin{array}{ccc}
0 & \sqrt{p} & 0 \\
0 & 0 & 0 \\
0 & 0 & 0%
\end{array}%
\right) ,\ \ E_{2}=\left(
\begin{array}{ccc}
0 & 0 & \sqrt{p} \\
0 & 0 & 0 \\
0 & 0 & 0%
\end{array}%
\right)  \label{E7}
\end{equation}%
and the single qutrit Kraus operators for the phase damping channel are
given as

\begin{equation}
E_{0}=\sqrt{1-p}\left(
\begin{array}{ccc}
1 & 0 & 0 \\
0 & 1 & 0 \\
0 & 0 & 1%
\end{array}%
\right) ,\ \ E_{1}=\sqrt{p}\left(
\begin{array}{ccc}
1 & 0 & 0 \\
0 & \omega  & 0 \\
0 & 0 & \omega ^{2}%
\end{array}%
\right) ,  \label{7}
\end{equation}%
The single qutrit Kraus operators for the depolarizing channel are given by
[31]

\begin{equation*}
E_{0}=\sqrt{1-p}I_{3},\ E_{1}=\sqrt{\frac{p}{8}}Y,\ E_{2}=\sqrt{\frac{p}{8}}%
Z,\ E_{3}=\sqrt{\frac{p}{8}}Y^{2},\ E_{4}=\sqrt{\frac{p}{8}}YZ
\end{equation*}

\begin{equation}
E_{5}=\sqrt{\frac{p}{8}}Y^{2}Z,\ E_{6}=\sqrt{\frac{p}{8}}YZ^{2},\ \ E_{7}=%
\sqrt{\frac{p}{8}}Y^{2}Z^{2},\ \ E_{8}=\sqrt{\frac{p}{8}}Z^{2}  \label{E8}
\end{equation}%
where $I_{3}$ is the identity matrix of order 3.

\begin{equation}
Y=\left(
\begin{array}{ccc}
0 & 1 & 0 \\
0 & 0 & 1 \\
1 & 0 & 0%
\end{array}%
\right) ,\ \ Z=\left(
\begin{array}{ccc}
1 & 0 & 0 \\
0 & \omega  & 0 \\
0 & 0 & \omega ^{2}%
\end{array}%
\right)   \label{9}
\end{equation}%
In the above equations, $p$ represents the quantum noise parameter and $%
\omega =e^{\frac{2\pi i}{3}}$. The evolution of the initial density matrix
of the composite system when it is influenced by local and multi-local
environments is given in Kraus operator form as
\begin{equation}
\rho _{f}=\sum\limits_{i,j,k}(E_{j}^{B}E_{k}^{A})\rho
_{AR}(E_{j}^{B}E_{k}^{A})^{\dagger }
\end{equation}%
and the evolution of the system when it is influenced by global environment
is given in Kraus operator representation as
\begin{equation}
\rho _{f}=\sum\limits_{i,j,k}(E_{i}^{AB}E_{j}^{B}E_{k}^{A})\rho
_{AR}(E_{i}^{AB}E_{j}^{B}E_{k}^{A})^{\dagger }
\end{equation}%
where $E_{k}^{A}=E_{m}^{A}\otimes I_{3},$ $I_{2}\otimes E_{j}^{B}$ are the
Kraus operators of the multilocal coupling of qubit and qutrit individually
and $E_{i}^{AB}=E_{m}^{A}\otimes E_{n}^{A}$ are the Kraus operators of the
collective coupling of the hybrid system. Using equations (5-10) along with
the initial density matrix of as given in equation (1) and taking the
partial transpose over the smaller subsystem (qubit), we find the
eigenvalues of the final density matrix. Let the decoherence parameters for
local and global noise of the qubit and qutrit be $p_{1}$, $p_{2}$ and $p$
respectively. The entanglement for all mixed states $\rho _{AB}$ of a
qubit-qutrit system is well quantified by the negativity [32]%
\begin{equation}
N(\rho _{AB})=\max \{0,\sum\limits_{k}\left\vert \lambda
_{k}^{T_{A}(-)}\right\vert \})
\end{equation}%
where $\lambda _{k}^{T_{A}(-)}$\ represents the negative eigenvalues of the
partial transpose of the density matrix $\rho _{AB}$ with respect to the
smaller subsystem. The eigenvalues of the partial transpose matrix when only
the qubit is influenced by the amplitude damping channel are given by
\begin{eqnarray}
\lambda _{1,2} &=&\frac{1}{2}\cos ^{2}r  \notag \\
\lambda _{3} &=&-\frac{1}{2}(-1+p_{1})\cos ^{2}r  \notag \\
\lambda _{4} &=&\frac{1}{2}(-1+p_{1})\cos ^{2}r  \notag \\
\lambda _{5} &=&-\frac{1}{2}(-1+p_{1})\sin ^{2}r  \notag \\
\lambda _{6} &=&\frac{1}{2}(-1+p_{1})\sin ^{2}r
\end{eqnarray}%
The only possible negative eigenvalue is the fourth one and the negativity
is calculated using the relation given in equation (11). The eigenvalues of
the partial transpose matrix when only qutrit is influenced by the amplitude
damping channel are given by%
\begin{eqnarray}
\lambda _{1} &=&-\frac{1}{2}(-1+p_{2})\cos ^{2}r  \notag \\
\lambda _{2,3} &=&\frac{1}{4}(p_{2}\mp \sqrt{p_{2}^{2}-4(-1+p_{2})\cos ^{4}r}%
)  \notag \\
\lambda _{4,5} &=&-\frac{1}{2}(-1+p_{2})\sin ^{2}r  \notag \\
\lambda _{6} &=&\frac{1}{2}(\cos ^{2}r+p_{2}\sin ^{2}r)
\end{eqnarray}%
The only possible negative eigenvalue is the second one. The eigenvalues of
the partial transpose matrix when both qubit and qutrit are influenced by
the amplitude damping channel are given by
\begin{eqnarray}
\lambda _{1} &=&-\frac{1}{2}(-1+p_{2})\cos ^{2}r  \notag \\
\lambda _{2} &=&\frac{1}{2}(-1+p_{1})(-1+p_{2})\sin ^{2}r  \notag \\
\lambda _{3} &=&-\frac{1}{2}(1+p_{1})(-1+p_{2})\sin ^{2}r  \notag \\
\lambda _{4} &=&-\frac{1}{2}(-1+p_{1})(\cos ^{2}r+p_{2}\sin ^{2}r)  \notag \\
\lambda _{5,6} &=&\left. \left(
\begin{array}{c}
\frac{1}{8}(p_{1}+2p_{2}+p_{1}p_{2}+p_{1}\cos (2r)-p_{1}p_{2}\cos (2r) \\
\mp 2\sqrt{%
\begin{array}{c}
4(-1+p_{1})(-1+p_{2})\cos ^{4}r+ \\
((p_{1}+p_{2})\cos ^{2}r+(1+p_{1})p_{2}\sin ^{2}r)^{2}%
\end{array}%
}]%
\end{array}%
\right) \right.
\end{eqnarray}%
The only possible negative eigenvalue is the fifth one. The eigenvalues of
the partial transpose matrix when the system is influenced by the global
noise of amplitude damping channel are given by
\begin{eqnarray}
\lambda _{1} &=&\frac{1}{2}(-1+p)(-1+p_{2})\cos ^{2}r  \notag \\
\lambda _{2} &=&\frac{1}{4}(-1+p_{1})^{2}(1+p+p_{2}-pp_{2}+(-1+p)(-1+p_{2})%
\cos (2r))  \notag \\
\lambda _{3} &=&\frac{1}{2}(-1+p)(-1+p_{1})^{2}(-1+p_{2})\sin ^{2}r  \notag
\\
\lambda _{4} &=&-\frac{1}{2}(-1+p)(-1-2p_{1}+p_{1}^{2})(-1+p_{2})\sin ^{2}r
\notag \\
\lambda _{5,6} &=&\left. \left(
\begin{array}{c}
\frac{1}{4}[p\cos ^{2}r+2p_{1}\cos ^{2}r-p_{1}^{2}\cos ^{2}r++p_{2}\cos
^{2}r-pp_{2}\cos ^{2}r \\
+p\sin ^{2}r+2pp_{1}\sin ^{2}r-pp_{1}^{2}\sin ^{2}r+p_{2}\sin
^{2}r-pp_{2}\sin ^{2}r \\
+2p_{1}p_{2}\sin ^{2}r-2pp_{1}p_{2}\sin ^{2}r-p_{1}^{2}p_{2}\sin
^{2}r+pp_{1}^{2}p_{2}\sin ^{2}r \\
\mp \sqrt{%
\begin{array}{c}
4(-1+p)(-1+p_{1})^{2}(-1+p_{2})\cos ^{4}r \\
+((-2p_{1}+p_{1}^{2}+p(-1+p_{2})-p_{2})\cos ^{2}r \\
-(-1-2p_{1}+p_{1}^{2})(p(-1+p_{2})-p_{2})\sin ^{2}r)^{2}%
\end{array}%
}]%
\end{array}%
\right) \right.
\end{eqnarray}%
The only possible negative eigenvalue is the fifth one. The eigenvalues of
the partial transpose matrix when only the qubit is influenced by the
depolarizing channel are given by%
\begin{eqnarray}
\lambda _{1,2} &=&\frac{1}{2}\sin ^{2}r  \notag \\
\lambda _{3} &=&\frac{1}{4}(-2\cos ^{2}r+3p_{1}\cos ^{2}r)  \notag \\
\lambda _{4,5,6} &=&-\frac{1}{4}(-2+p_{1})\cos ^{2}r
\end{eqnarray}%
The only possible negative eigenvalue is the third one. The eigenvalues of
the partial transpose matrix when only qutrit is influenced by the
depolarizing channel are given by%
\begin{eqnarray}
\lambda _{1,2,3} &=&\frac{1}{32}(8-3p_{2}+8\cos (2r)-9p_{2}\cos (2r))  \notag
\\
\lambda _{4} &=&\frac{1}{32}(-8+15p_{2}-8\cos (2r)+9p_{2}\cos (2r))  \notag
\\
\lambda _{5,6} &=&\frac{1}{32}(8-3p_{2}+(-8+9p_{2})\cos (2r))
\end{eqnarray}%
The only possible negative eigenvalue is the fourth one. The eigenvalues of
the partial transpose matrix when both the qubit and the qutrit are
influenced by the depolarizing channel are given by%
\begin{eqnarray}
\lambda _{1,2} &=&\frac{1}{32}(8-3p_{2}+(-8+9p_{2})\cos (2r))  \notag \\
\lambda _{3,4} &=&\frac{1}{32}((16-12p_{2}+p_{1}(-8+9p_{2}))\cos
^{2}r+6p_{2}\sin ^{2}r)  \notag \\
\lambda _{5,6} &=&\left. \left(
\begin{array}{c}
\frac{1}{64}(8p_{1}+12p_{2}-9p_{1}p_{2}\mp 2\sqrt{2}\sqrt{%
2(-1+p_{1})^{2}(8-9p_{2})^{2}\cos ^{4}r} \\
+8p_{1}\cos (2r)-9p_{1}p_{2}\cos (2r))%
\end{array}%
\right) \right.
\end{eqnarray}%
The only possible negative eigenvalue is the fifth one. The eigenvalues of
the partial transpose matrix when the system is influenced by the global
noise of depolarizing channel are given by%
\begin{eqnarray}
\lambda _{1,2} &=&\left. \left(
\begin{array}{c}
\frac{1}{512}%
(96p+128p_{1}-144pp_{1}-64p_{1}^{2}+72pp_{1}^{2}+96p_{2}-108pp_{2}-144p_{1}p_{2}
\\
+162pp_{1}p_{2}+72p_{1}^{2}p_{2}-81pp_{1}^{2}p_{2}\mp 2\sqrt{2}\sqrt{%
2(8-9p)^{2}(-1+p_{1})^{4}(8-9p_{2})^{2}\cos ^{4}r} \\
+128p_{1}\cos (2r)-144pp_{1}\cos (2r)-64p_{1}^{2}\cos (2r)+72pp_{1}^{2}\cos
(2r) \\
-144p_{1}p_{2}\cos (2r)+162pp_{1}p_{2}\cos (2r)+72p_{1}^{2}p_{2}\cos
(2r)-81pp_{1}^{2}p_{2}\cos (2r))%
\end{array}%
\right) \right.   \notag \\
\lambda _{3,4} &=&\frac{1}{256}(64-24p_{2}+3p(-8+9p_{2})-(-8+9p)(-8+9p_{2})%
\cos (2r))  \notag \\
\lambda _{5,6} &=&\left. \left(
\begin{array}{c}
\frac{1}{256}%
((3p(4-6p_{1}+3p_{1}^{2})(-8+9p_{2})-8(4(-4+3p_{2})-2p_{1}(-8+9p_{2}) \\
+p_{1}^{2}(-8+9p_{2})))\cos ^{2}r+6(p(8-9p_{2})+8p_{2})\sin ^{2}r)%
\end{array}%
\right) \right.
\end{eqnarray}%
The only possible negative eigenvalue is the first one. The eigenvalues of
the partial transpose matrix when only qubit is influenced by the phase
damping channel are given by%
\begin{eqnarray}
\lambda _{1,2} &=&\frac{1}{2}\cos ^{2}r  \notag \\
\lambda _{3} &=&-\frac{1}{2}\sqrt{\cos ^{4}r-p_{1}\cos ^{4}r}  \notag \\
\lambda _{4} &=&\frac{1}{2}\sqrt{\cos ^{4}r-p_{1}\cos ^{4}r}  \notag \\
\lambda _{5,6} &=&\frac{1}{2}\sin ^{2}r
\end{eqnarray}%
The only possible negative eigenvalue is the third one. The eigenvalues of
the partial transpose matrix when only qutrit is influenced by the phase
damping channel are given by%
\begin{eqnarray}
\lambda _{1,2} &=&\frac{1}{2}\cos ^{2}r  \notag \\
\lambda _{3} &=&-\frac{1}{2}\sqrt{(1-3p_{2}+3p_{2}^{2})\cos ^{4}r}  \notag \\
\lambda _{4} &=&\frac{1}{2}\sqrt{(1-3p_{2}+3p_{2}^{2})\cos ^{4}r}  \notag \\
\lambda _{5,6} &=&\frac{1}{2}\sin ^{2}r
\end{eqnarray}%
The only possible negative eigenvalue is the third one. The eigenvalues of
the partial transpose matrix when both qubit and qutrit are influenced by
the phase damping channel are given by%
\begin{eqnarray}
\lambda _{1,2} &=&\frac{1}{2}\cos ^{2}r  \notag \\
\lambda _{3} &=&-\frac{1}{2}\sqrt{(-1+p_{1})(-1+3p_{2}-3p_{2}^{2})\cos ^{4}r}
\notag \\
\lambda _{4} &=&\frac{1}{2}\sqrt{(-1+p_{1})(-1+3p_{2}-3p_{2}^{2})\cos ^{4}r}
\notag \\
\lambda _{5,6} &=&\frac{1}{2}\sin ^{2}r
\end{eqnarray}%
The only possible negative eigenvalue is the third one. The eigenvalues of
the partial transpose matrix when the system is influenced by the global
noise of phase damping channel are given by%
\begin{eqnarray}
\lambda _{1,2} &=&\frac{1}{2}\cos ^{2}r  \notag \\
\lambda _{3} &=&-\frac{1}{2}\sqrt{%
(1-3p+3p^{2})(-1+p_{1})^{2}(1-3p_{2}+3p_{2}^{2})\cos ^{4}r}  \notag \\
\lambda _{4} &=&\frac{1}{2}\sqrt{%
(1-3p+3p^{2})(-1+p_{1})^{2}(1-3p_{2}+3p_{2}^{2})\cos ^{4}r}  \notag \\
\lambda _{5,6} &=&\frac{1}{2}\sin ^{2}r
\end{eqnarray}%
The only possible negative eigenvalue is the third one. The negativity is
calculated using equation (5) for all the above cases and results are
discussed in detail in the next section.

\section{Discussions}

In this work, we investigate the effect of decoherence on a qubit-qutrit
system in non-inertial frames. In figure 1, we plot negativity as a function
of Rob's acceleration, $r$\ for decoherence parameters $p_{i}=0.2$ for
amplitude damping channel. It is seen that maximal entanglement degradation
occurs under global noise. It is also seen that the entanglement is degraded
heavily as we increase the value of Rob's acceleration. However, the
entanglement loss is consistent for all the cases and no ESD behaviour is
seen for any acceleration. In figure 2, we plot the negativity as a function
of Rob's acceleration, $r$\ for $p_{1}=p_{2}=0.2$ (multi-local noise) and $%
p=0.2$ (global noise) for depolarizing channel. It is seen that depolarizing
channel heavily degrades the entanglement as compared to amplitude damping
channel, particularly in case of global decoherence. In figure 3, we plot
the negativity as a function of Rob's acceleration, $r$\ for $%
p_{1}=p_{2}=0.2 $ (multi-local noise) and $p=0.2$ (global noise) for phase
damping channel. A similar behaviour of amplitude damping and phase damping
channels is seen towards entanglement degradation.

In figure 4, we plot the negativity as a function of Rob's acceleration, $r$%
\ for $p_{1}=p_{2}=p=0.2$ for amplitude damping, depolarizing and phase
damping channels. It is shown that depolarizing channel influences the
entanglement of the system more heavily as compared to the other two
channels. On the other hand, for higher values of decoherence parameters,
the amplitude damping channel have more influence on the entanglement
degradation as clear from figure 5. Further more, it is also seen that no
ESD occurs for any acceleration of Rob for the entire range of decoherence
parameters.

\section{Conclusions}

We analyze the effect of decoherence on a qubit-qutrit system under the
influence of decoherence in non-inertial frames. We consider different noise
models such as amplitude damping, depolarizing and phase damping channels
with different couplings of the system and the environment. We show that the
entanglement sudden death can be avoided in non-inertial frames. However,
degradation of entanglement is seen due to Unruh effect. It is seen that for
lower values of decoherence parameters, the depolarizing channel heavily
degrades the entanglement of the system as compared to the amplitude damping
and phase damping channels. However, for higher values of decoherence
parameters, amplitude damping channel heavily degrades the entanglement of
the system. In conclusion, no ESD occurs for any value of Rob's acceleration.

{\huge Figures captions}\newline
\textbf{Figure 1}. The negativity is plotted as a function of Rob's
acceleration, $r$\ for $p_{1}=p_{2}=0.2$ (multi-local noise) and $p=0.2$
(global noise) for amplitude damping channel.\newline
\textbf{Figure 2}. The negativity is plotted as a function of Rob's
acceleration, $r$\ for $p_{1}=p_{2}=0.2$ (multi-local noise) and $p=0.2$
(global noise) for depolarizing channel.\newline
\textbf{Figure 3}. The negativity is plotted as a function of Rob's
acceleration, $r$\ for $p_{1}=p_{2}=0.2$ (multi-local noise) and $p=0.2$
(global noise) for phase damping channel.\newline
\textbf{Figure 4}. The negativity is plotted as a function of Rob's
acceleration, $r$\ for $p_{1}=p_{2}=p=0.2$ for amplitude damping,
depolarizing and phase damping channels.\newline
\textbf{Figure 5}. The negativity is plotted as a function of Rob's
acceleration, $r$\ for $p_{1}=p_{2}=p=0.5$ for amplitude damping,
depolarizing and phase damping channels.\newline
{\Huge Table Caption}\newline
\textbf{Table 1}. Single qubit Kraus operators for amplitude damping,
depolarizing and phase damping channels where $p$ represents the decoherence
parameter.\newline

\begin{figure}[tbp]
\begin{center}
\vspace{-2cm} \includegraphics[scale=0.8]{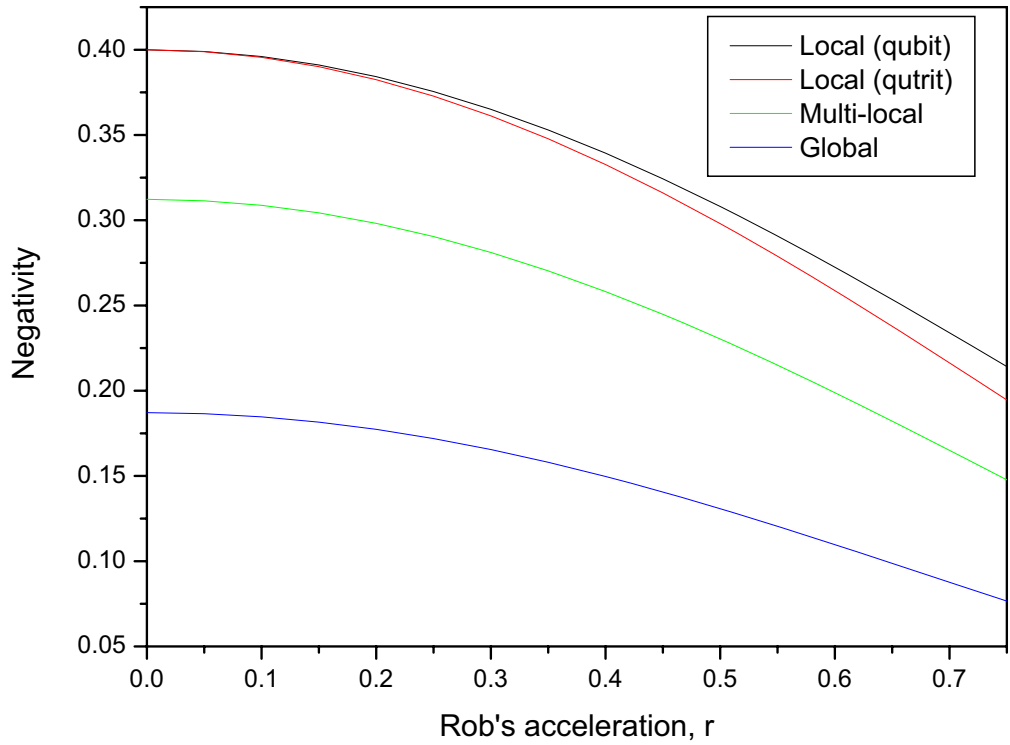} \\[0pt]
\end{center}
\caption{The negativity is plotted as a function of Rob's acceleration, $r$\
for $p_{1}=p_{2}=0.2$ (multi-local noise) and $p=0.2$ (global noise) for
amplitude damping channel.}
\end{figure}

\begin{figure}[tbp]
\begin{center}
\vspace{-2cm} \includegraphics[scale=0.8]{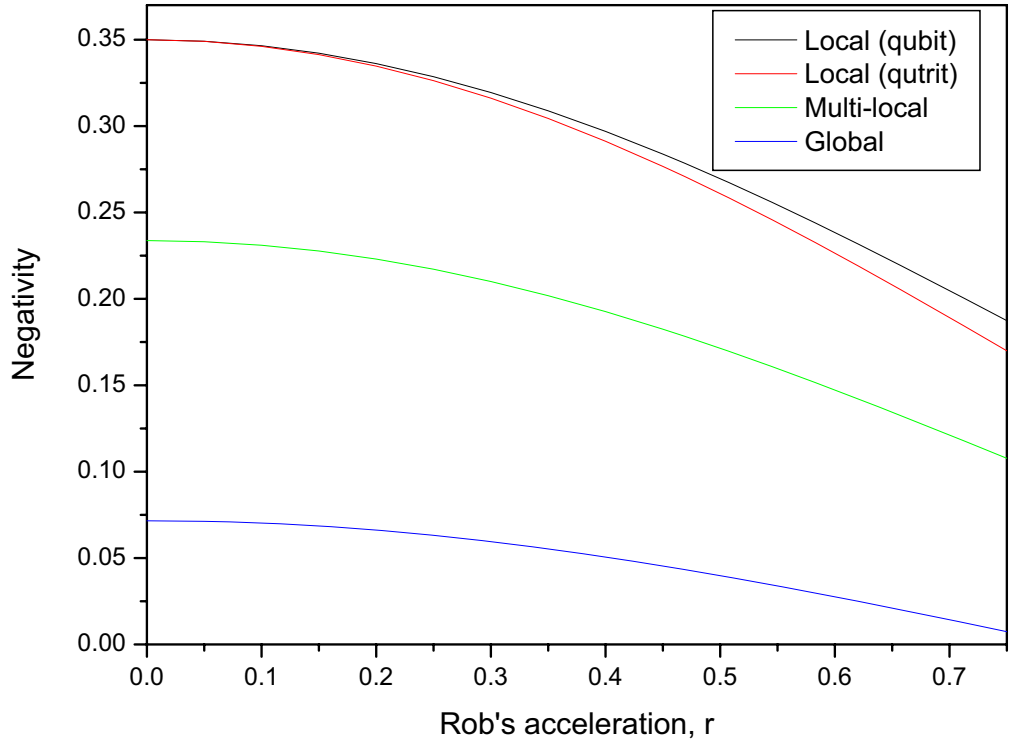} \\[0pt]
\end{center}
\caption{The negativity is plotted as a function of Rob's acceleration, $r$\
for $p_{1}=p_{2}=0.2$ (multi-local noise) and $p=0.2$ (global noise) for
depolarizing channel.}
\end{figure}

\begin{figure}[tbp]
\begin{center}
\vspace{-2cm} \includegraphics[scale=0.8]{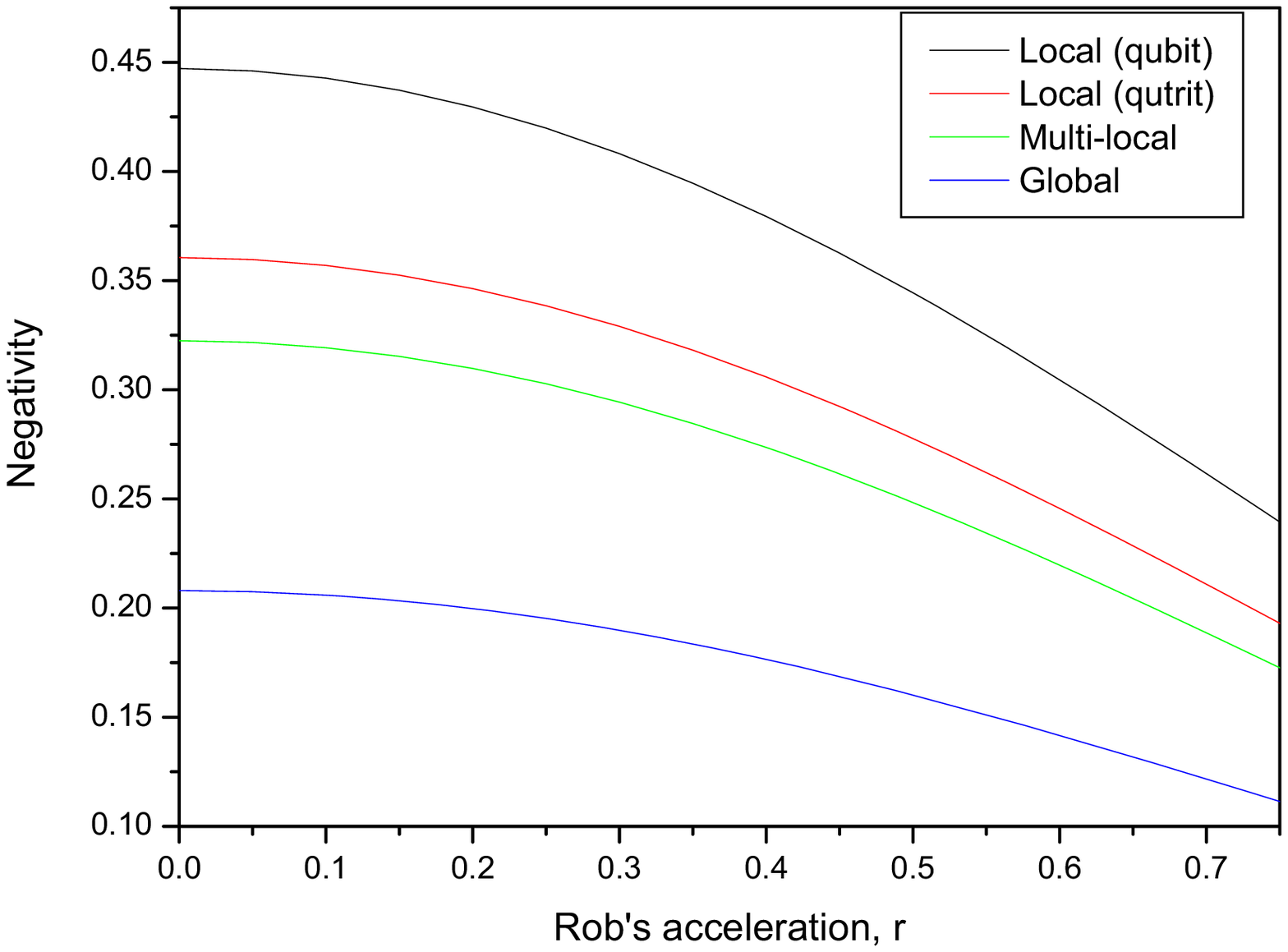} \\[0pt]
\end{center}
\caption{The negativity is plotted as a function of Rob's acceleration, $r$\
for $p_{1}=p_{2}=0.2$ (multi-local noise) and $p=0.2$ (global noise) for
phase damping channel.}
\end{figure}

\begin{figure}[tbp]
\begin{center}
\vspace{-2cm} \includegraphics[scale=0.8]{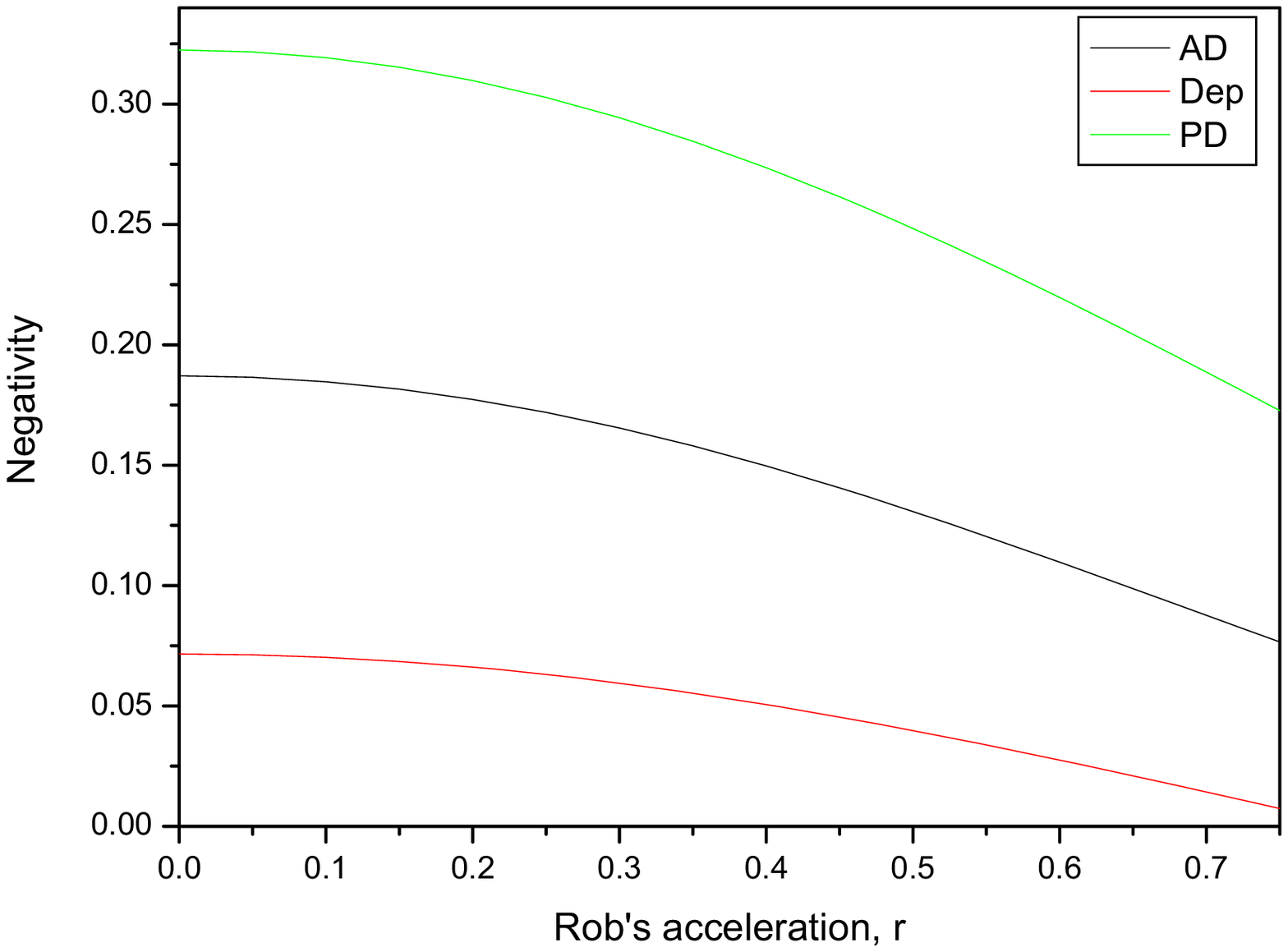} \\[0pt]
\end{center}
\caption{The negativity is plotted as a function of Rob's acceleration, $r$\
for $p_{1}=p_{2}=p=0.2$ for amplitude damping, depolarizing and phase
damping channels.}
\end{figure}

\begin{figure}[tbp]
\begin{center}
\vspace{-2cm} \includegraphics[scale=0.8]{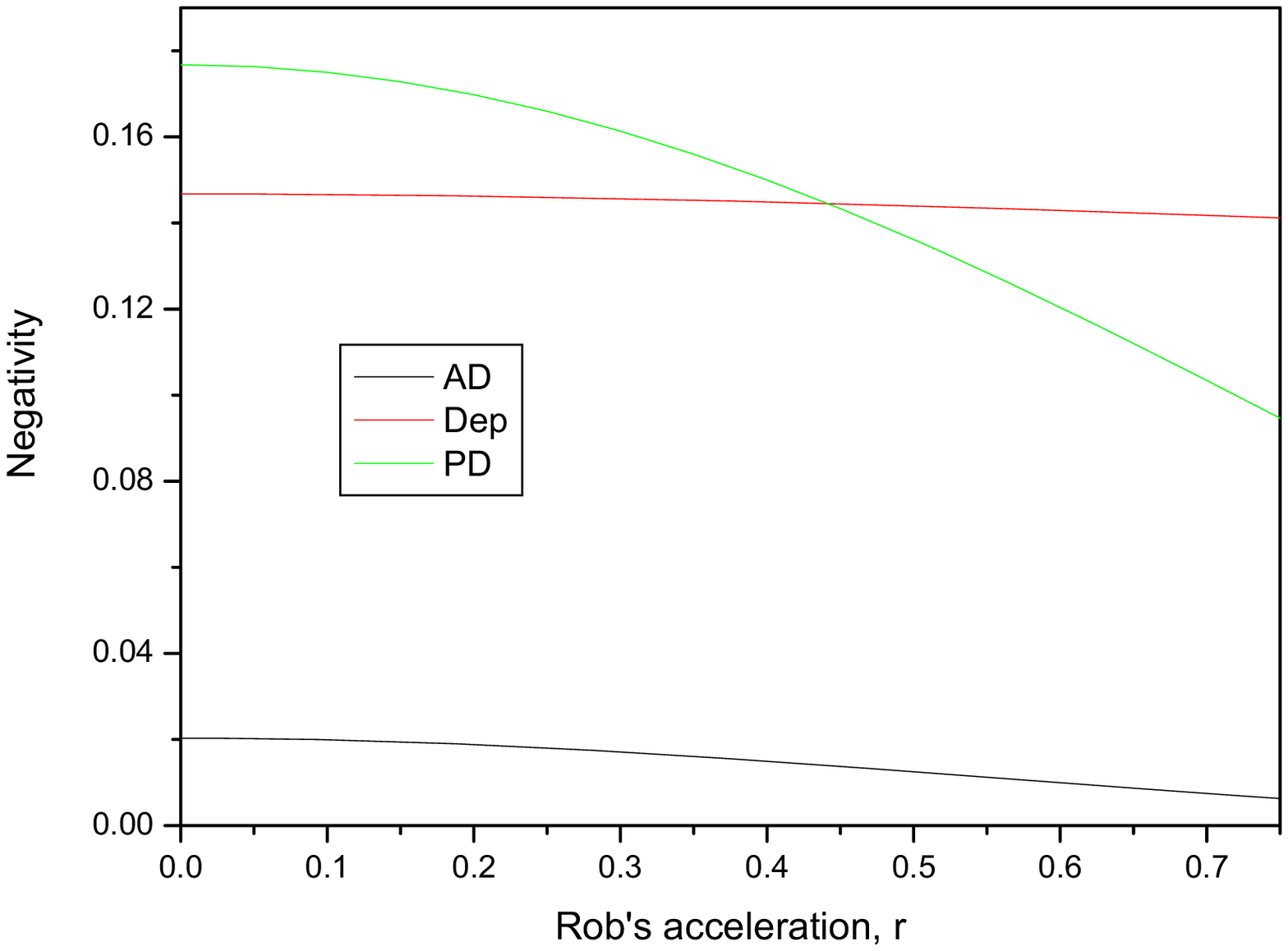} \\[0pt]
\end{center}
\caption{The negativity is plotted as a function of Rob's acceleration, $r$\
for $p_{1}=p_{2}=p=0.5$ for amplitude damping, depolarizing and phase
damping channels.}
\end{figure}

\begin{table}[tbh]
\caption{Single qubit Kraus operators for amplitude damping, depolarizing
and phase damping channels where $p$ represents the decoherence parameter.}
\label{di-fit}$%
\begin{tabular}{|l|l|}
\hline
&  \\
$\text{Depolarizing channel}$ & $%
\begin{tabular}{l}
$E_{0}=\sqrt{1-\frac{3p}{4}I},\quad E_{1}=\sqrt{\frac{p}{4}}\sigma _{x}$ \\
$E_{2}=\sqrt{\frac{p}{4}}\sigma _{y},\quad \quad $\ $\ E_{3}=\sqrt{\frac{p}{4%
}}\sigma _{z}$%
\end{tabular}%
$ \\
&  \\ \hline
&  \\
$\text{Amplitude damping channel}$ & $E_{0}=\left[
\begin{array}{cc}
1 & 0 \\
0 & \sqrt{1-p}%
\end{array}%
\right] ,$ $E_{1}=\left[
\begin{array}{cc}
0 & \sqrt{p} \\
0 & 0%
\end{array}%
\right] $ \\
&  \\ \hline
&  \\
$\text{Phase damping channel}$ & $E_{0}=\left[
\begin{array}{cc}
1 & 0 \\
0 & \sqrt{1-p}%
\end{array}%
\right] ,E_{1}=\left[
\begin{array}{cc}
0 & 0 \\
0 & \sqrt{p}%
\end{array}%
\right] $ \\
&  \\ \hline
\end{tabular}%
$%
\end{table}

\end{document}